\documentclass[11pt]{article}
\usepackage{array}
\usepackage{epsf} 
\usepackage{amsmath,color}
\usepackage{amssymb}
\usepackage{epsfig}
\usepackage{latexsym}
\usepackage{amsfonts}
\usepackage{dcolumn}
\usepackage{varioref}
\usepackage{ifthen}
\usepackage{graphicx}
\usepackage{amssymb,amsmath,amsthm}
\usepackage{slashed}
\begin{document}
\parindent 0mm 
\setlength{\parskip}{\baselineskip} 
\thispagestyle{empty}
\pagenumbering{arabic} 
\setcounter{page}{1}
\begin{center}
{\Large {\bf Electromagnetic proton form factors in dual large-$N_c$ QCD: an update}}
\end{center}

\begin{center}
{\bf B. Bisschoff} $^{(a)}$, {\bf C. A. Dominguez} $^{(a)}$,
{\bf L. A. Hernandez} $^{(a)}$\\
\vspace{.2cm}

{\it $^{(a)}$ Centre for Theoretical and Mathematical Physics, 
	and Department of Physics, University of
	Cape Town, Rondebosch 7700, South Africa}\\
\end{center}

\begin{abstract}
\noindent
An updated determination is presented of the electric and magnetic form factors of the proton, in the framework of a dual-model realization of QCD in the limit of an infinite number of colours. Very good agreement with data is obtained in the space-like region up to $ q^2 \simeq - \,30 \,{\mbox{GeV}}^2$. In particular, the ratio $\mu_P \, G_E(q^2) / G_M(q^2)$ is predicted in very good agreement with recoil polarization measurements from Jefferson Lab, up to $q^2 \simeq - 8.5 \,{\mbox{GeV}}^2$.
\end{abstract}

\noindent
It has been established long ago that QCD for a large number of colours \cite{tHooft}, $N_c \rightarrow \infty$, i.e. ${\mbox{QCD}}_{\infty}$, leads to a very simple hadronic spectrum, i.e. an infinite number of zero-width resonances \cite{Witten}. The masses and couplings of these states, though, remain unspecified so that  models are required to fix these parameters. Since in the real world $N_c = 3$, a naive estimate of the corrections to ${\mbox{QCD}}_{\infty}$ would be at the level of 30\%. However, in practice this could be an overestimate. For instance, while finite width corrections in ${\mbox{QCD}}_{\infty}$ are of order $1/N_c$, in the hadronic sector they are only of
${\cal{O}}(\Gamma_R/M_R)$, where $\Gamma_R$ and $M_R$ are the width and the mass of a resonance, respectively. In the case of the $\rho$-meson this is below 20\%. Furthermore,  ${\mbox{QCD}}_{\infty}$ models of hadron form factors in the space-like region 
are expected to be largely insensitive to finite-width effects in the time-like region beyond threshold. An independent argument suggesting that corrections to ${\mbox{QCD}}_{\infty}$ are more likely to be at the 10\%  level, rather than 30\%, may be found in \cite{Coleman}.
A few models of the ${\mbox{QCD}}_{\infty}$ spectrum have been proposed in the heavy-quark sector \cite{HQ1,HQ2}, as well as in the light-quark sector \cite{LQ}.\\
The peculiar hadronic spectrum of  ${\mbox{QCD}}_{\infty}$ is reminiscent of the dual-resonance model of Veneziano \cite{Veneziano1,Veneziano2}, the precursor of string theory. In the case of three-point functions, this observation motivated a specific model for the masses and couplings, $\mbox{Dual-QCD}_{\infty}$, leading to an Euler's beta function of the Veneziano-type \cite{pionFF}. It should be mentioned that models of this type precede QCD. In fact, they were first proposed for the electromagnetic form factors of the proton \cite{EM10,EM11}, the $\Delta(1236)$ \cite{Delta1}, and for the radiative decays of mesons \cite{CADRD}. They were also used in purely hadronic processes \cite{Bryan1,Bryan2}, and $SU(2)\times SU(2)$ chiral symmetry breaking corrections \cite{GTR1,GTR2}. Ultimately, the basic idea of a tower of radial excitations, of e.g. the $\rho$-meson, can be traced back to the extension of Sakurai's Vector Meson Dominance Model (VMD)\cite{Sakurai}, to Extended Vector Meson Dominance \cite{EVMD1,EVMD2,EVMD3}.\\

In this paper we update a previous determination of the proton form factors in the framework of $\mbox{Dual-QCD}_{\infty}$ \cite{CADTT} in order to account for new experimental data  over an extended range of four-momentum squared, in the space-like region. In particular, new data on the ratio $\mu_P \, G_E(q^2) / G_M(q^2)$. It should be recalled that the empirical historical assumption of this ratio to be approximately constant was found in serious conflict with Jefferson Lab polarization transfer data \cite{JL11,JL12} up to $  q^2 \simeq  - 6 \,{\mbox{GeV}}^2$. Indeed, the ratio $\mu_p \, G_E(q^2) / G_M(q^2)$ was found to be a monotonically decreasing function of $q^2$. More recent data at higher values of $q^2$, up to $q^2 = - 8.5 \, {\mbox{GeV}^2}$, shows a continuation of this trend. \\

A generic electromagnetic form factor in the framework of $\mbox{QCD}_{\infty}$ is given by
\begin{equation}
F(s) = \sum_{n=0}^{\infty}\, \frac{C_n}{(M_n^2 - s)}\,,\label{F}
\end{equation}
where $s$ is the four-momentum squared, and the masses of the (zero-width) vector-meson resonances, $M_n$, as well as their couplings $C_n$, are unspecified and in need of a specific model. In dual-$\mbox{QCD}_{\infty}$ these parameters are fixed by requiring  the form factors to be given by an Euler's beta function, i.e.
\begin{equation}
C_n = \frac{\Gamma(\beta - 1/2)}{\alpha' \sqrt{\pi}}
\; \frac{(-1)^n}{\Gamma(n+1)} \; \frac{1}{\Gamma(\beta -1 - n)} \,,
\label{Cn}
\end{equation}
where $\beta$ is a free parameter determining the asymptotic behaviour of the form factor in the space-like region ($s < 0$), and $\alpha' = 1/ ( 2 \, M_\rho^2)$ is the universal string tension entering the $\rho$-meson Regge trajectory
\begin{equation}
\alpha_\rho(s) \, = \, 1 \, + \, \alpha' \,(s - M_\rho^2)\,.\label{alpha}
\end{equation}
The masses of the radial excitations are given by \cite{EVMD1,EVMD2,EVMD3}
\begin{equation}
M_n^2 = M_\rho^2 (1 + 2n). \label{mass}
\end{equation}
To compare with data, this mass formula gives for the first three radial excitations $M_{\rho'} \simeq 1340 \,{\mbox{MeV}}$,
 $M_{\rho''} \simeq 1720 \,{\mbox{MeV}}$,
and $M_{\rho'''} \simeq 2034 \,{\mbox{MeV}}$, in reasonable agreement with the experimental  values
\cite{PDG} $M_{\rho'} = 1465 \,{\mbox{MeV}}$,  $M_{\rho''} = 1720 \,{\mbox{MeV}}$,
and $M_{\rho'''} = 2000-2200 \,{\mbox{MeV}}$, all with very large widths, $\Gamma \simeq 200 - 400\,{\mbox{MeV}}$. Instead of the linear mass formula, Eq.(\ref{mass}), non-linear forms may be required to match the asymptotic Regge behaviour to the Operator Product Expansion (OPE) of current correlators at short distances \cite{OPE}. However, the differences in the values of the masses for the first few states is only at the level of a few percent. Given that the contribution to the form factor from high mass states is factorial suppressed by the beta function, these differences have no impact in the predictions.\\
Substituting Eqs.(\ref{Cn}) and (\ref{mass}) into Eq.(\ref{F}) gives the dual-$QCD_{\infty}$ form factor
\begin{eqnarray} 
F(s) \, &=& \, \frac{1}{\sqrt{\pi}} \; \frac{\Gamma(\beta - 1/2)}{\Gamma(\beta - 1)}\; B\left(  \beta-1, \frac{1}{2} - \alpha' \, s  \right) \nonumber  \\ [.3cm]
&=& \frac{\Gamma(\beta - 1/2)}{\sqrt{\pi}} \,\sum_{n=0}^{\infty}
\frac{(-1)^n}{\Gamma(n+1)}\, \frac{1}{\Gamma(\beta - 1 - n)} \, \frac{1}{[n+1 - \alpha_\rho(s)]} \,, \label{FF}
\end{eqnarray}
where $B(x,y) = \Gamma(x)\, \Gamma(y)/\Gamma(x+y)$ is the Euler Beta function. This form factor is analytic in the space-like region ($s < 0$), while it has an infinite number of poles for time-like $s$, i.e. $s > 0$, and non-integer values of $\beta$. For integer $\beta$ the number of poles is finite, but obviously there is no discontinuous behaviour.  In fact, its imaginary part is given by
\begin{equation}
{\mbox{Im}}\, F(s) = \frac{\Gamma(\beta - 1/2)}{\alpha' \,\sqrt{\pi}}
\sum_{n=0}^{\infty} \frac{(-1)^n}{\Gamma(n+1)} \, \frac{1}{\Gamma(\beta - 1 - n)} \; \pi \, \delta(M_n^2 \, - \, s)\,. \label{ImF}
\end{equation}
The numerical value of the single free parameter, $\beta$, can be determined, e.g. from a fit to data in the space-like region, or from data for the root-mean-squared radius. It has been shown  for the pion electromagnetic form factor \cite{pionFF}  that both methods produce consistent results. In the case of $\beta=2$, $F(s)$ has only one pole, and thus it reduces to ordinary VMD.\\
\begin{figure}
\begin{center}
\includegraphics[width=6.2cm,height=5.2cm]{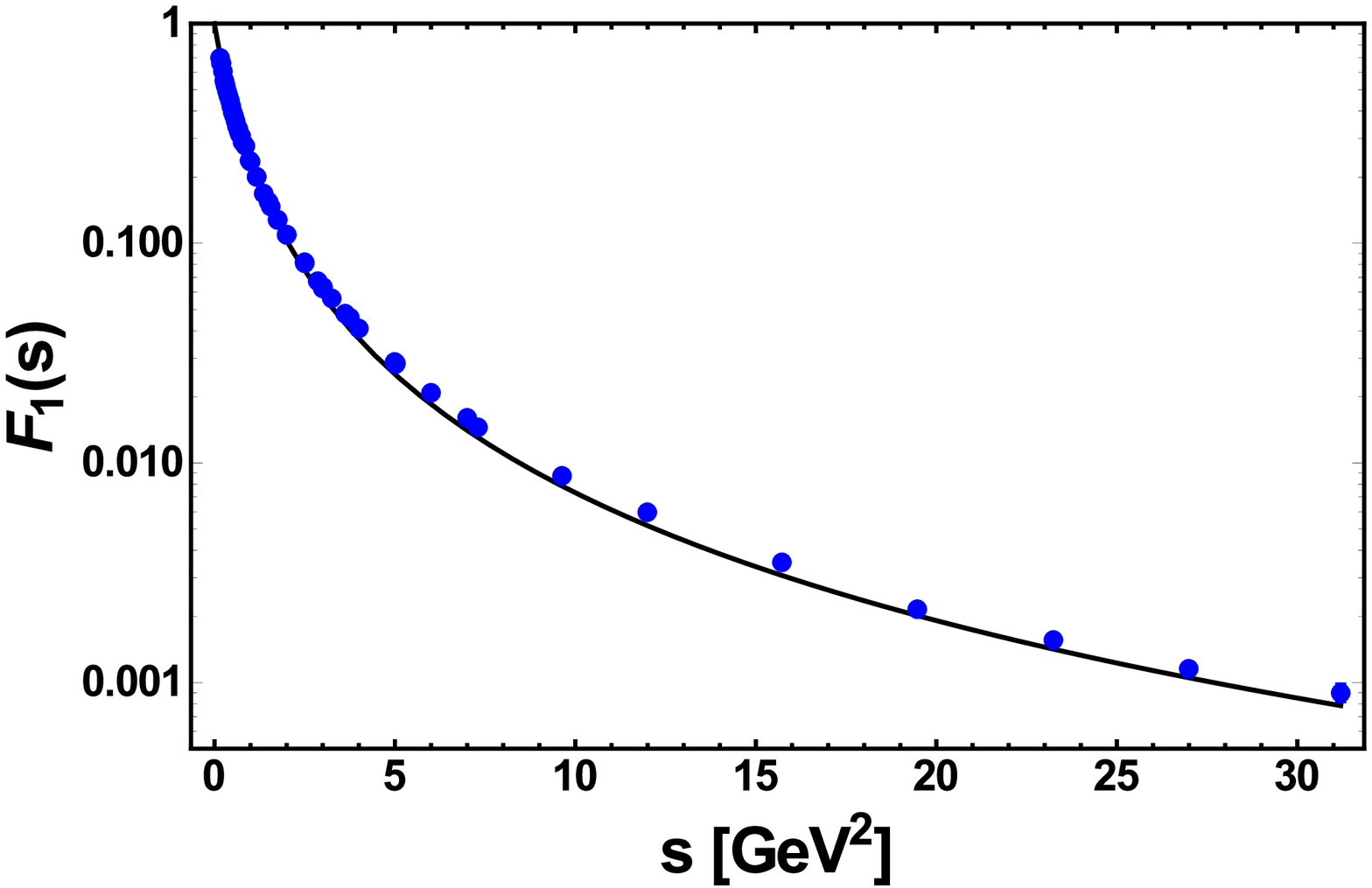}
\includegraphics[width=6.2cm,height=5.2cm]{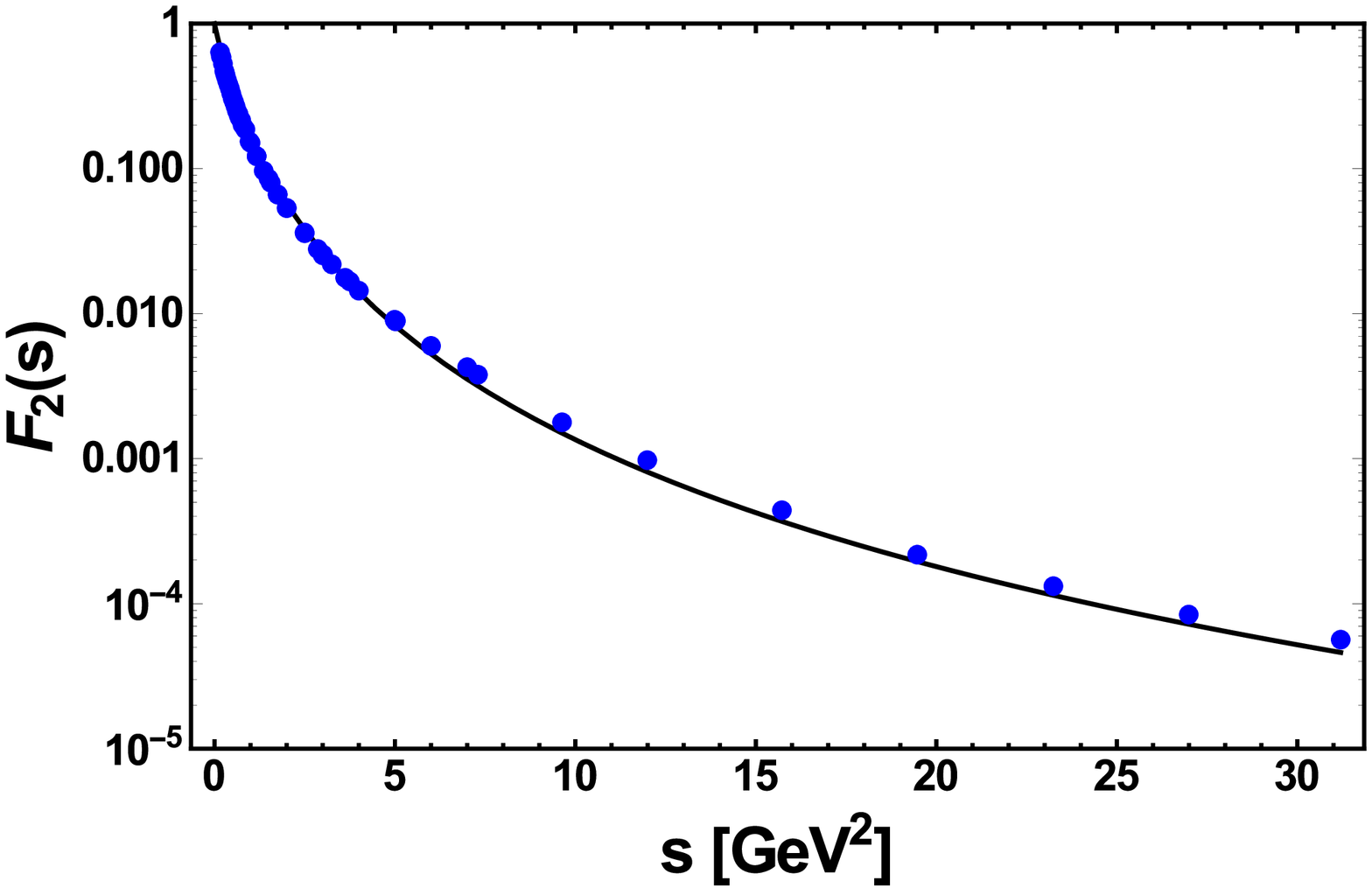}
\caption{\small The form factors $F_1(s)$ (left panel), and $F_2(s)$ (right panel) in the space-like region as a function of $s=-q^2$. Dots are the experimental points obtained from inverting Eqs. (8) and (9), and using the data base compilation from \cite{Brash} . Solid line is the prediction of the $\mbox{Dual-QCD}_{\infty}$ expression, Eq.(\ref{FF}), with $\beta_1= 3.105$, and $\beta_2= 4.305$. Notice the logarithmic scale.}
\label{fig:figure1}
\end{center}
\end{figure}
Turning to the electromagnetic form factors of the proton, the Dirac and Pauli form factors, $F_1(q^2)$, and $F_2(q^2)$ respectively, are defined as  
\begin{equation}
\langle N(p_2)| V_\mu^{EM}(0)|N(p_1\rangle = \bar{u}_N(p_2) \left[F_1(q^2) \, \gamma_\mu + \frac{i \, \kappa}{2\, M_N} \, F_2(q^2) \, \sigma_{\mu\nu}\, q^\nu  \right] u_N(p_1)\,, \label{F12}
\end{equation}
where $N$ stands for the proton, $q^2 = (p_2 - p_1)^2$, $\kappa \equiv \mu_N -1$ and $M_N$ are the proton's magnetic moment and mass, respectively, with $F_{1,2}(0) = 1$. These form factors have no kinematical singularities and  satisfy dispersion relations, so that the expression Eq.(\ref{F}) applies to them. In relation to electron-proton elastic scattering experiments another set of form factors, the Sach's form factors $G_E(q^2)$
and $G_M(q^2)$ are convenient, and defined as
\begin{equation}
G_E(q^2) = F_1(q^2) - \kappa \, \tau \, F_2(q^2)\,, \label{GE}
\end{equation}
\begin{equation}
G_M(q^2) = F_1(q^2) + \kappa \,  F_2(q^2)\,, \label{GM}
\end{equation}
where $\tau \equiv -q^2/4M_p^2$, and the normalization is $G_E(0) = 1$, $G_M(0) = \mu_p$.\\
\begin{figure}[ht!]
\begin{center}
\includegraphics[width=6.2cm,height=5.2cm]{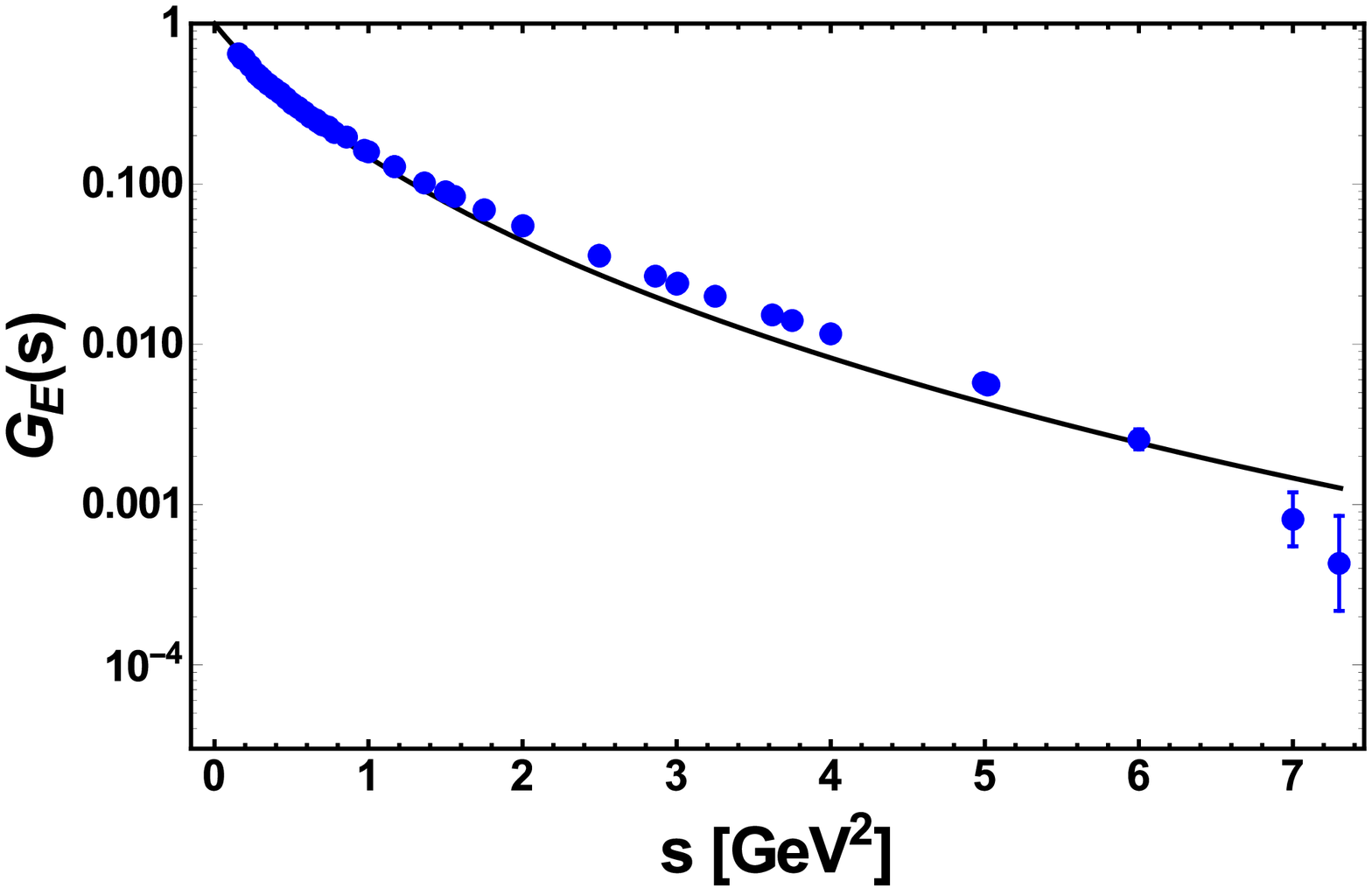}
\includegraphics[width=6.2cm,height=5.2cm]{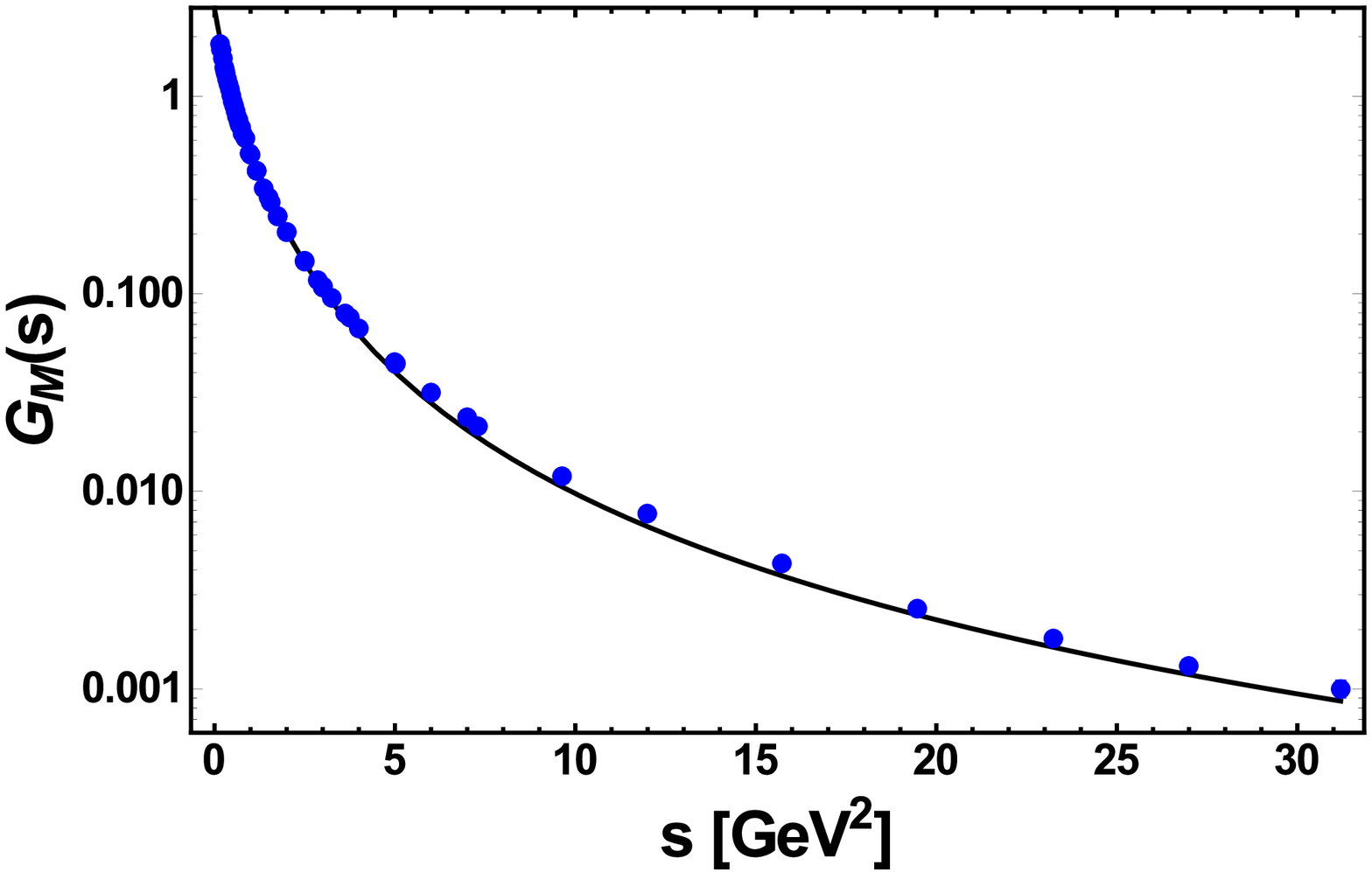}
\caption{\small The form factors $G_E(s)$ (left panel), and $G_M(s)$ (right panel) in the space-like region as a function of $s=-q^2$. Dots are the experimental points using the data base compilation from \cite{Brash}. Solid curves are obtained from Eqs.(\ref{GE}) and (\ref{GM}), with $F_{1,2}(s)$ as in Fig.1. Notice the logarithmic scale.}
\label{fig:figure2}
\end{center}
\end{figure}
Concerning the experimental data, we shall use the compilation of reanalyzed  world data for $G_E(q^2)$ and $G_M(q^2)$ by Brash et al. \cite{Brash}, as well as more  recent polarization transfer data from Jefferson Lab for  $\mu_p \, G_E(q^2) / G_M(q^2)$ \cite{JL11,JL12} . In order to find the optimal values of the free parameters $\beta_1$ and $\beta_2$, we determine  initial values of $F_1(q^2)$ and $F_2(q^2)$ for an initial set $\beta_{1,2}$, leading to corresponding values of $G_E(q^2)$, $G_M(q^2)$, which are compared to the data. The process is iterated until a best fit to the latter is obtained. In this fashion we find
\begin{equation}
\beta_1 = 3.105 \,,
\end{equation}
\begin{equation}
\beta_2 = 4.305 \,.
\end{equation}
\begin{figure}[ht!]
\begin{center}
\includegraphics[width=7.0cm, height=8.0cm]{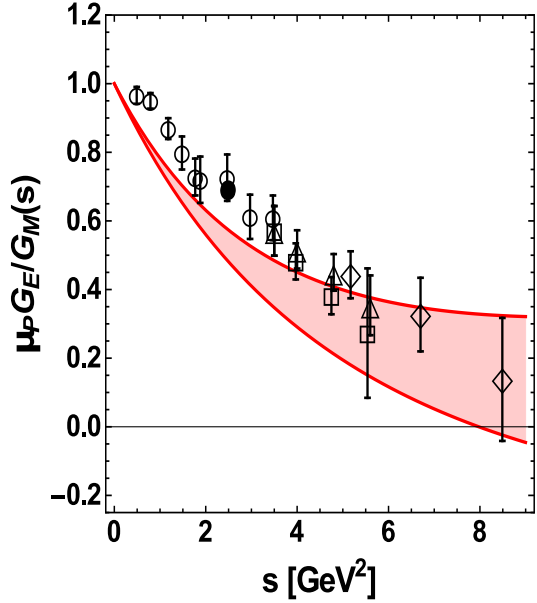}
\caption{\small The prediction for the form factor ratio as a function of $s = - q^2$, together with polarization transfer data from Jefferson Lab \cite{JL11,JL12,JL21,JL22} , using results for $G_{E,M}(s)$ as in Fig.2. The band corresponds to keeping the parameter $\beta_1$ fixed, and allowing $\beta_2$ to change in the interval $\beta_2 = 4.225 - 4.385$. This change has a much smaller impact on the form factors themselves.}
\end{center}
\end{figure}
Results for the form factors $F_{1,2}(s)$ are shown in Fig. 1, together with the data determined from  that of $G_{E,M}(s)$. The electric and magnetic form factors in $\mbox{Dual-QCD}_{\infty}$, $G_{E,M}(s)$, are shown in Fig.2.  Having thus determined $\beta_{1,2}$, the ratio $\mu_p \, G_E(s) / G_M(s)$ becomes a prediction, shown in Figure 3. The band corresponds to keeping $\beta_1$ fixed, and allowing $\beta_2$ to move in the range $\beta_2 = 4.225-4.385$.
Changing $\beta_2$ in this range has very little impact on results for the form factors themselves.

Finally, we consider the various electromagnetic radii, starting with those associated with $F_{1,2}(q^2)$. Differentiating Eq.(\ref{FF}) with respect to $q^2$, at $q^2=0$, gives 
\begin{equation}
\langle r^2_{1,2}\rangle = 6 \,\alpha' \, \left[ \psi\left( \beta_{1,2} - \frac{1}{2}\right) - \psi\left( \frac{1}{2}\right) \right] \,,
\end{equation}
where $\psi(x)$ is the digamma function. The resulting electric and magnetic radii, associated with $G_E$ and $GM$, respectively, are
\begin{equation}
\langle r_M^2 \rangle^{1/2} \, \simeq \, \langle r_E^2 \rangle^{1/2} \simeq\, 0.8 \,{\mbox{fm}} \,, \label{radii}
\end{equation}
in  agreement with current values \cite{radius1,radius2,radius3}. It should be emphasized that the above values are the result of a one-parameter fit to each of the two form factors over a very large range of (space-like) momentum transfer. With the radius being sensitive to very low $q^2$ data, the results above provide additional strong support for the $\mbox{Dual-QCD}_{\infty}$ model. If one were to restrict the form factor fit to very low values of $q^2$, any increase over the result in Eq.(\ref{radii}) would be at the expense of the agreement between the form factors and data at high $q^2$. 

\section*{Acknowledgement}
This work was supported in part by the National Research Foundation (South Africa).  LAH acknowledges funding assistance from the University of Cape Town and the National Research Foundation (South Africa).

\end{document}